\documentclass[a4paper]{article}

\usepackage{INTERSPEECH2020}
\usepackage{multirow}
\usepackage{stmaryrd}
\usepackage{hyperref}
\usepackage{subcaption}
\usepackage{xcolor}
\usepackage{cite}
\usepackage[multiple]{footmisc}
\usepackage{numprint}
\npthousandsep{,}

\title{LibriMix: An Open-Source Dataset for Generalizable Speech Separation}

\name{Joris Cosentino$^1$, Manuel Pariente$^1$, Samuele Cornell$^2$, Antoine Deleforge$^1$, Emmanuel Vincent$^1$
\thanks{Experiments presented in this paper were carried out using the Grid'5000 testbed, supported by a scientific interest group hosted by Inria and including CNRS, RENATER and several Universities as well as other organizations (see \url{https://www.grid5000.fr}).}}
\address{
  $^1$Universit\'e de Lorraine, CNRS, Inria, LORIA, F-54000 Nancy, France
  \newline
  $^2$Department of Information Engineering, Universit\`a Politecnica delle Marche, Italy}
\email{joris.cosentino@inria.fr, manuel.pariente@inria.fr, s.cornell@pm.univpm.it, antoine.deleforge@inria.fr, emmanuel.vincent@inria.fr}

\begin{document}

\maketitle

\begin{abstract}
In recent years, wsj0-2mix has become the reference dataset for single-channel speech separation.
Most deep learning-based speech separation models today are benchmarked on it.
However, recent studies have shown important performance drops when models trained on wsj0-2mix are evaluated on other, similar datasets.
To address this generalization issue, we created LibriMix, an open-source alternative to wsj0-2mix, and to its noisy extension, WHAM!.
Based on LibriSpeech, LibriMix consists of two- or three-speaker mixtures combined with ambient noise samples from WHAM!.
Using Conv-TasNet, we achieve competitive performance on all LibriMix versions. 
In order to fairly evaluate across datasets, we introduce a third test set based on VCTK for speech and WHAM! for noise.
Our experiments show that the generalization error is smaller for models trained with LibriMix than with WHAM!, in both clean and noisy conditions.
Aiming towards evaluation in more realistic, conversation-like scenarios, we also release a sparsely overlapping version of LibriMix's test set.
\end{abstract}

\noindent\textbf{Index Terms}: Speech separation, generalization, corpora.

\section{Introduction}
A fundamental problem towards robust speech processing in real-world acoustic environments is to be able to automatically extract or separate target source signals present in an input mixture recording \cite{BookEVincent}.
To date, state-of-the-art performance on the single-channel speech separation task is achieved by deep learning based models \cite{DPCLHershey2016, LSTMLuo2018, ConvTasnetLuo2018, DPRNNLuo2020, Wavesplit2020Zeghidour}. In particular, end-to-end models which directly process the time-domain samples seem to obtain the best performance \cite{CompehensiveBahmaninezhad2019, HeitDemyst2019}.
Such systems (e.g. Conv-TasNet \cite{ConvTasnetLuo2018}, Dual-path RNN \cite{DPRNNLuo2020} or Wavesplit \cite{Wavesplit2020Zeghidour}) perform so well in separating fully overlapping speech mixtures from the wsj0-2mix dataset \cite{DPCLHershey2016} that the separated speech estimates are almost indistinguishable from the reference signals. This led to the development of WHAM!\cite{Wichern2019WHAM} and WHAMR!\cite{Maciejewski2020WHAMR}, respectively the noisy and reverberant extensions of wsj0-2mix.

While these datasets have moved the field towards more realistic and challenging scenarios, there are still steps to be made. In fact, a recent study 
reports important drops of performance when Conv-TasNet is trained on wsj0-2mix and tested on other comparable datasets \cite{EmpiricalDolby2020}.
This suggests that, even though Conv-TasNet's separation quality is close to perfect on wsj0-2mix, the ability to generalize to speech coming from a wider range of speakers and recorded in slightly different conditions has not yet been achieved. 
Additionally, fully overlapping speech mixtures such as the ones from wsj0-2mix are unnatural. Real-world overlap ratios are typically in the order of 20\% or less in natural meetings  \cite{etin2006AnalysisOO} and casual dinner parties \cite{barker2018Chime}. A few studies have shown that speech separation algorithms trained on fully overlapping speech mixtures do not generalize well to such sparsely overlapping mixtures \cite{SparseMenne2019, Wavesplit2020Zeghidour}. Finally, models relying on some kind of speaker identity representation \cite{DPCLHershey2016, Wavesplit2020Zeghidour, DeepCASA2019} cannot easily detect overfitting, since wsj0-2mix's speakers are shared between the training and validation sets.

There have been few initiatives to address these issues. A sparsely overlapping version of wsj0-2mix proposed in \cite{SparseMenne2019} has shown the limitation of Deep Clustering \cite{DPCLHershey2016} on such mixtures. As the original utterances are the same as the ones from wsj0-2mix, we expect the generalization issue to remain the same.
In \cite{EmpiricalDolby2020}, a new speech separation dataset based on LibriTTS \cite{LibriTTS2019} has been designed. The results show that generalizability is improved thanks to the variability of recording conditions and the larger number of unique speakers in the dataset. Sadly, the dataset is limited to two-speaker mixtures without noise, and has not been open-sourced. LibriCSS \cite{LibriCSSChen_2020}, an open-source dataset for sparsely overlapping continuous speech separation, has recently been released. While it addresses most of the shortcomings of wsj0-2mix, its short 10-hour duration restricts its usage to evaluation rather than training purposes. Real diner-party recordings  \cite{barker2018Chime, segbroeck2019dipco} as well as meeting recordings \cite{ISCI2013, AMI2006} are also available. While these are natural recordings, the clean speech signals for individual sources are not available\footnote{Close-talk signals are available as references, but these are too corrupted for the evaluation of modern separation algorithms.} and thus, speech separation algorithms cannot be directly evaluated in terms of usual speech separation metrics \cite{SDRVincent2006, SISDRLeroux2019}.

In this work, we introduce LibriMix, an open-source dataset for generalizable noisy speech separation composed of two- or three-speaker mixtures, with or without noise. The speech utterances are taken from LibriSpeech \cite{panayotov2015librispeech} and the noise samples from WHAM!\cite{Wichern2019WHAM}.
An additional test set based on VCTK \cite{VCTK} is designed for fair cross-dataset evaluation. 
We evaluate the generalization ability of Conv-TasNet when trained on LibriMix or WHAM! and show that LibriMix leads to better generalization in both clean and noisy conditions. 
Stepping further towards real-world scenarios, we introduce a sparsely overlapping version of LibriMix's test set with varying amount of overlap. The scripts used to generate these datasets are publicly released\footnote{\url{https://github.com/JorisCos/LibriMix}}\footnote{\url{https://github.com/JorisCos/VCTK-2Mix}}\footnote{\url{https://github.com/popcornell/SparseLibriMix}}.

The paper is organised as follows. We explain LibriMix's design and give some insights about its characteristics in Section \ref{sec:librimix_dataset}.
In Section \ref{sec:results}, we report experimental results on LibriMix as well as across datasets. We conclude in Section \ref{sec:conclusions}.

\section{Datasets} \label{sec:librimix_dataset}
In the following, we present existing speech separation datasets derived from Wall Street Journal (WSJ0), and introduce our new datasets derived from LibriSpeech. Statistics about the original speech datasets and the speech separation datasets derived from them can be found in Tables \ref{tab:initial_datasets} and \ref{tab:resulting_datasets}, respectively.

\begin{table}[h]
\centering
\caption{Statistics of original speech datasets.}
\label{tab:initial_datasets}
\begin{tabular}{c|c|ccc}
\hline
Dataset & Split & Hours & \begin{tabular}[c]{@{}c@{}}per-spk \\ minutes\end{tabular} & \begin{tabular}[c]{@{}c@{}} \# Speakers\end{tabular} \\ \hline
\multirow{3}{*}{WSJ0}                                                          & si\_tr\_s  & 25    & 15 & 101 \\ 
                                                                               & si\_dt\_05 & 1.5   & 11 & 8   \\ 
                                                                               & si\_et\_05 & 2.3   & 14 & 10  \\ \hline
\multirow{4}{*}{\begin{tabular}[c]{@{}c@{}}LibriSpeech\\ clean\end{tabular}} & train-360  & 364 & 25 & 921 \\

                                                                               & train-100  & 101 & 25 & 251 \\ 
                                                            
                                                                               & dev        & 5.4   & 8  & 40  \\ 
                                                                               & test       & 5.4   & 8  & 40  \\ \hline
VCTK                                                                           & test       & 44    & 24 & 109 \\ \hline
\end{tabular}
\end{table}

\begin{table}[h]
\centering
\caption{Statistics of derived speech separation datasets.}
\label{tab:resulting_datasets}
\begin{tabular}{c|c|ccc}
\hline
Dataset & Split & \begin{tabular}[c]{@{}c@{}} \# Utterances\end{tabular} & Hours \\ \hline
\multirow{3}{*}{wsj0-\{2,3\}mix}                                  & train     & \numprint{20000} & 30   \\ 
                                                            & dev        & \numprint{5000}  & 8    \\ 
                                                            & test      & \numprint{3000}  & 5   \\ \hline
\multirow{4}{*}{Libri2Mix}                                  & train-360 & \numprint{50800} & 212  \\  
                                                            & train-100 & \numprint{13900} & 58   \\  
                                                            & dev       & \numprint{3000}  & 11   \\  
                                                            & test      & \numprint{3000}  & 11     \\ \hline
\multirow{4}{*}{Libri3Mix}                                 & train-360 & \numprint{33900} & 146 \\  
                                                            & train-100 & \numprint{9300}  & 40   \\  
                                                             & dev       & \numprint{3000}    & 11    \\  
                                                            & test      & \numprint{3000}    & 11     \\ \hline
\begin{tabular}[c]{@{}c@{}}SparseLibri2Mix\end{tabular} & test      & \numprint{3000}      & 6      \\ \hline
\begin{tabular}[c]{@{}c@{}}SparseLibri3Mix\end{tabular} & test      & \numprint{3000}       &  6      \\ \hline
VCTK-2mix                                                   & test      & \numprint{3000}  & 9  \\ \hline
\end{tabular}
\end{table}

\subsection{WSJ0, wsj0-2mix and WHAM!}
The WSJ0 dataset was designed in 1992 as a new corpus for automatic speech recognition (ASR)\cite{WSJ_CSR}.
It consists of read speech from the Wall Street Journal. 
It was recorded at 16 kHz using a close-talk Sennheiser HMD414 microphone.
The wsj0-2mix dataset \cite{DPCLHershey2016} uses three subsets of WSJ0: \texttt{si\_tr\_s},  \texttt{si\_dt\_05}  and \texttt{si\_et\_05} which all come from the 5k vocabulary part of WSJ0. This represents around 30~h of speech from 119 speakers. 
Table \ref{tab:initial_datasets} reports details on speaker and hour distributions within the subsets.

The wsj0-2mix datatet is made of a training set, a validation set and a test set.
The training and validation sets share common speakers from the \texttt{si\_tr\_s} subset and the test set is made from a combination of \texttt{si\_dt\_05} and \texttt{si\_et\_05}.
Speech mixtures are generated by mixing pairs of utterances from different speakers at random signal-to-noise ratios (SNRs). The SNR is drawn uniformly between 0 and 5 dB. 
Four variations of the dataset are available, which correspond to two different sampling rates (16~kHz and 8~kHz) and two modes (\textit{min} and \textit{max}).
In the \textit{min} mode, the mixture stops with the shortest utterance. 
In the \textit{max} mode, the shortest utterance is padded to the longest one.
The wsj0-2mix equivalent for three-speaker mixtures is called wsj0-3mix and was generated in a similar way \cite{DPCLHershey2016}. 
Note that, in order to generate more mixtures, utterances from WSJ0 were used multiple times in the three subsets.
Each utterance is repeated up to fifteen times, with an average of four times.

In the WHAM! dataset, wsj0-2mix was extended to include noisy speech mixtures. 
Noise samples recorded in coffee shops, restaurants, and bars were added to the mixtures so that the SNR between the loudest speaker and the noise varies from -6 to +3 dB.
The dataset follows the same structure as wsj0-2mix, with the same four variations and the three same subsets. In addition to separation in clean (\textit{sep\_clean}) and noisy conditions (\textit{sep\_noisy}), other enhancement tasks can be considered.
Statistics on noise durations can be seen in Table \ref{tab:Wham's_noises_time}.

WHAM! noises have been released under the CC BY-NC 4.0 License, but WSJ0 and derived data are proprietary (LDC). Note that no noisy version of wsj0-3mix has been released. 

\begin{table}[h]
\centering
\caption{Statistics of WHAM!'s noises.}
\label{tab:Wham's_noises_time}
\begin{tabular}{c|c|cc}
\hline
Datasets                                                               & Split & Hours & \begin{tabular}[c]{@{}c@{}}Number of \\ utterances \end{tabular} \\ \hline
\multirow{3}{*}{\begin{tabular}[c]{@{}c@{}}WHAM!\\ noise\end{tabular}} & train    & 58    & \numprint{20000}                                                             \\ 
 & dev & 14.7 & \numprint{5000} \\ 
 & test & 9    & \numprint{3000}\\ \hline
\end{tabular}
\end{table}

\subsection{LibriSpeech, LibriMix and sparse LibriMix}
LibriSpeech \cite{panayotov2015librispeech} is a read ASR corpus based on LibriVox audiobooks\footnote{https://librivox.org/}.
To avoid background noise in the reference signals, we only use the \texttt{train-clean-100}, \texttt{train-clean-360}, \texttt{dev-clean}, and \texttt{test-clean} subsets of LibriSpeech.
This represents around 470~h of speech from \numprint{1252} speakers, with a 60k vocabulary.
More statistics are given in Table \ref{tab:resulting_datasets}.

We propose a new collection of datasets derived from LibriSpeech and WHAM!'s noises which we call LibriMix. These datasets are entirely open source.

The two main datasets, Libri2Mix and Libri3Mix, consist of clean and noisy, two- and three-speaker mixtures.
Libri2Mix follows the exact same structure as WHAM! and allows for the same tasks.
Mirroring the organization of LibriSpeech, they have two training sets (\texttt{train-100}, \texttt{train-360}), one validation set (\texttt{dev}) and one test set (\texttt{test}).
In order to cover the \texttt{train-360} subset of LibriSpeech without repetition, training noise samples were speed-perturbed with factors of 0.8 and 1.2 as described in \cite{ko2015augmentation}.
Instead of relying on signal power to scale individual utterances as in wsj0-2mix, we rely on loudness units relative to full scale (LUFS) \cite{loudness}\footnote{Available at https://github.com/csteinmetz1/pyloudnorm}, expressed in dB. Based on the ITU-R BS.1770-4 recommendation \cite{loudness}, LUFS measure the perceived loudness of an audio signal. Compared to classical SNRs, LUFS better correlate with human perception, are silence-invariant, and are little sensitive to downsampling.

Speech mixtures are generated by randomly selecting utterances for different speakers. 
The loudness of each utterance is uniformly sampled between -25 and -33 LUFS.
Random noise samples with uniformly distributed loudness between -38 and -30 LUFS are then added to the speech mixtures. 
The noisy mixtures are then clipped to 0.9, if need be.
The resulting SNRs are normally distributed with a mean of 0~dB and a standard deviation of 4.1~dB in the clean condition and a mean of -2~dB and a standard deviation of 3.6~dB in the noisy condition.

Note that in \texttt{train-100} and \texttt{train-360} each utterance is only used once. 
For \texttt{dev} and \texttt{test}, 
the same procedure is repeated enough times to reach \numprint{3000} mixtures.
This results in around 280~h of noisy speech mixtures, against 45~h for WHAM!.
The variety of speakers is much wider in LibriMix's training set with around \numprint{1000} distinct speakers against 100 in WHAM!. The total number of unique words is also much larger, with 60k unique words in LibriMix against 5k in wsj0-2mix.

Stepping towards more realistic, conversation-like scenarios, we also release sparsely overlapping versions of LibriMix's two- and three-speaker test sets. We refer to these datasets as SparseLibri2Mix and SparseLibri3Mix. 
For each  mixture, we first sample speaker identities, then, for each speaker, we select an utterance from \texttt{test-clean}.
Cycling through the selected utterances, we keep adding sub-utterances whose boundaries were obtained with the Montreal Forced Aligner (MFA) \cite{mcauliffe2017montreal}, until a maximum length of 15~s has been reached. This mixing process ensures that each speaker utters semantically meaningful speech, which is important for future ASR experiments. 
We used the same loudness distribution as the non-sparse version but we sampled it for each sub-utterance. This allows for alternating dominant speakers in the mixtures \cite{Wavesplit2020Zeghidour}. 

For both two- and three-speaker versions, we produced 500 mixtures for six different amounts of speech overlap: 0\%, 20\%, 40\%, 60\% 80\%, and 100\%. For three-speaker mixtures we count the amount of three-speaker overlap and not the total overlap, which is higher because two-speaker overlap also occurs. 
Note that these overlap ratios reflect the amount of overlap of each sub-utterance with the preceding ones. Because sub-utterances don't have the same length, the real overlap ratios of the mixtures are lower, as it happens with \textit{max} versions of LibriMix and WHAM!. 

\begin{figure}[h]
\begin{subfigure}{\linewidth}
\centering
\hspace{-0.5cm}
\includegraphics{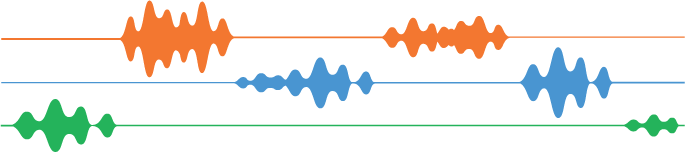}
\caption{}
\label{fig:sub3}
\end{subfigure}
\begin{subfigure}{.5\linewidth}
\centering
\includegraphics{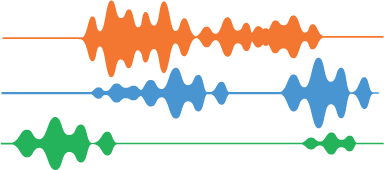}
\caption{}
\label{fig:sub1}
\end{subfigure}%
\begin{subfigure}{.5\linewidth}
\centering
\includegraphics{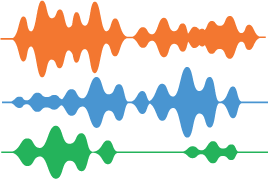}
\caption{}
\label{fig:sub2}
\end{subfigure}\\[1ex]
\caption{SparseLibri3Mix example with different 3-speaker overlap percentages: 
 (a) 0\% overlap, (b) 20\% overlap, (c) 100\% overlap.}
\label{fig:sparselibri2}
\end{figure}

Because WHAM! noise samples are short on average, the maximum mixture length was restricted to 15~s in order to obtain a reasonable number of samples for testing. 
Examples of such sparsely overlapping utterances can be visualized in Fig. \ref{fig:sparselibri2}.

\subsection{VCTK and VCTK-2mix}
We also release VCTK-2mix, an unmatched open-source test set derived from VCTK \cite{VCTK}. VCTK comprises 109 native English speakers reading newspapers. 
As VCTK utterances contain a significant amount of silence, we use energy-based voice activity detection to remove silent portions with a 20~dB threshold.

The mixing procedure for VCTK-2mix is identical to that for LibriMix.
The noise samples are also taken from WHAM!'s test set.
The resulting dataset contains around 9~h of speech with \numprint{3000} utterances from 108 speakers.

\section{Results}\label{sec:results}
In order to assess the results achievable using our newly released LibriMix datasets, we use the optimal configuration of Conv-TasNet reported in \cite{ConvTasnetLuo2018} for the separation tasks, as implemented in Asteroid  \cite{Asteroid2020} \footnote{\href{https://github.com/mpariente/asteroid}{github.com/mpariente/asteroid}}.
Training is done by maximizing the permutation-invariant, scale-invariant signal-to-distortion ratio (SI-SDR) \cite{PITYu2016, SDRVincent2006} on 3~s segments with a batch size of 24 and Adam \cite{Adam} as the optimizer. All the experiments are performed with the exact same parameters. Since the SI-SDR is undefined for silent sources, results reported on all \textit{max} versions correspond to models trained on the corresponding \textit{min} version.

\subsection{Results on LibriMix}
The results achieved by Conv-TasNet on the clean and noisy versions of Libri2mix and Libri3Mix are reported in Table \ref{tab:Baselinebis} and compared with the Ideal Binary Mask (IBM) and the Ideal Ratio Mask (IRM) for a short time Fourier transform (STFT) window size of 32~ms. Conv-TasNet was trained on \texttt{train-360}, which leads to better performance than \texttt{train-100}. Results are reported in terms of SI-SDR improvement compared to the input mixture (SI-SDR$_\text{i}$).
We refer to the clean two-speaker separation task as \textit{2spk-C}, to the noisy one as \textit{2spk-N}, etc. We see that for two-speaker mixtures, Conv-TasNet outperforms ideal masks in clean conditions and is on par with them in noisy conditions, as in \cite{ConvTasnetLuo2018, FilterbankDesign2019Pariente}. However, oracle performance is still out of reach for three-speaker mixtures, with and without noise.

\begin{table}[h]
\centering
\caption{SI-SDR$_\text{i}$ (dB) achieved on LibriMix (SI-SDR for the "Input" column).}
\label{tab:Baselinebis}
\begin{tabular}{c|c|cccc}
\hline
                            & mode                         & Input & IRM & IBM & Conv-TasNet \\ \hline
\multirow{2}{*}{2spk-C} & 8k min                       & 0.0   &  12.9   & 13.7    & 14.7            \\
                            & 16k max                  & 0.0   &  14.1   & 14.5    & 16            \\ \hline
\multirow{2}{*}{2spk-N} & 8k min                       & -2.0   & 12    & 12.6    & 12            \\
                        & \multicolumn{1}{l|}{16k max} & -2.8  & 13.4    &  13.7  &  13.5           \\ \hline
\multirow{2}{*}{3spk-C} & 8k min                       & -3.4      & 13.1    & 13.9    & 12.1           \\
                        & \multicolumn{1}{l|}{16k max} &  -3.7     &   14.5 &  14.9   &  13           \\ \hline
\multirow{2}{*}{3spk-N} & 8k min                       &  -4.4     &  12.6   & 13.3    &    10.4\\
                        & 16k max                      & -5.2      &  14.1  & 14.4    &    10.9       \\ \hline
\end{tabular}
\end{table}

\subsection{Results on SparseLibriMix}
We report the results obtained on the 8 kHz test sets of SparseLibri2Mix and SparseLibri3Mix in Table \ref{tab:sparse_results}, in clean and noisy conditions. We used the same 8 kHz models as in Table \ref{tab:Baselinebis}, which were trained on non-sparse LibriMix.
It can be seen that, for both two- and three-speaker mixtures, the higher the overlap, the lower the SI-SDR$_\text{i}$, as was also shown in \cite{Wavesplit2020Zeghidour}. In the 100\% overlap case we obtain results similar to the ones in Table \ref{tab:Baselinebis} for the non-sparse, 8kHz \textit{min} version. The values are slightly higher here because mixtures are not truncated to the shortest utterance. Interestingly, we see that Conv-TasNet performs \textit{worse} than IRM for smaller overlaps. This suggests that there is still room for improvement for source separation of  sparsely-overlapping mixtures.

\begin{table}[t]
\centering	
\setlength\tabcolsep{3pt}
\small
\caption{SI-SDR$_\text{i}$ (dB) achieved on SparseLibriMix (8kHz). Conv-TasNet is abreviated TCN.}
\begin{tabular}{c|cc|cc|cc|cc}
\hline
           & \multicolumn{2}{c|}{2spk-C} & \multicolumn{2}{c|}{2spk-N} & \multicolumn{2}{c|}{3spk-C} & \multicolumn{2}{c}{3spk-N} \\ \hline
Overlap & IRM          & TCN          & IRM          & TCN          & IRM          & TCN          & IRM          & TCN          \\
0\%        & 43.7         & 31.9         & 16.1         & 14.5         & 44.2         & 24.8         & 18.7         & 13.0         \\
20\%       & 19.6         & 20.0         & 14.7         & 13.9         & 18.1         & 15.8         & 15.6         & 12.1             \\
40\%       & 16.2         & 17.6         & 13.8         & 13.2         & 16.4         & 14.4         & 14.9         & 11.7             \\
60\%       & 14.9         &  16.3        & 13.3         & 12.7         & 15.5         & 13.8         & 14.4         & 11.5             \\
80\%       & 14.2         & 15.7         & 13           & 12.5         & 14.6         & 13.1         & 13.9         & 11             \\
100\%      & 13.8         & 15.3         & 12.7         & 12.2         & 14.3         & 12.5         & 13.6         & 10.7      \\
\hline
\end{tabular}
\label{tab:sparse_results}
\end{table}

\subsection{Dataset comparisons}
\begin{figure}[h]
    \centering
    \includegraphics[width=8cm]{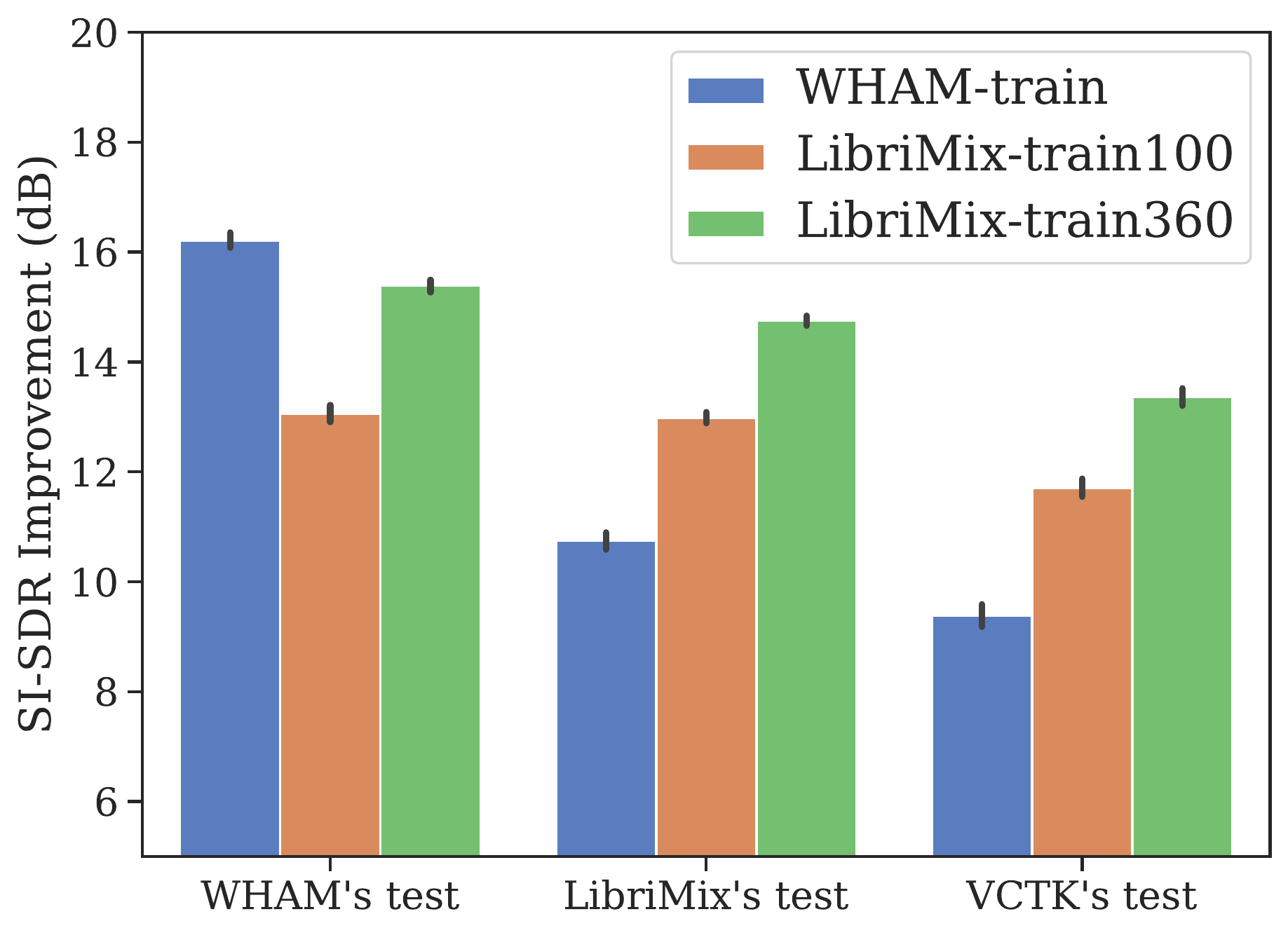}
    \caption{Cross-dataset evaluation on the clean separation task. Errors bars indicate 95\% confidence intervals.}
    \label{fig:sep_clean}
\end{figure}

\begin{figure}[t]
    \centering
    \includegraphics[width=8cm]{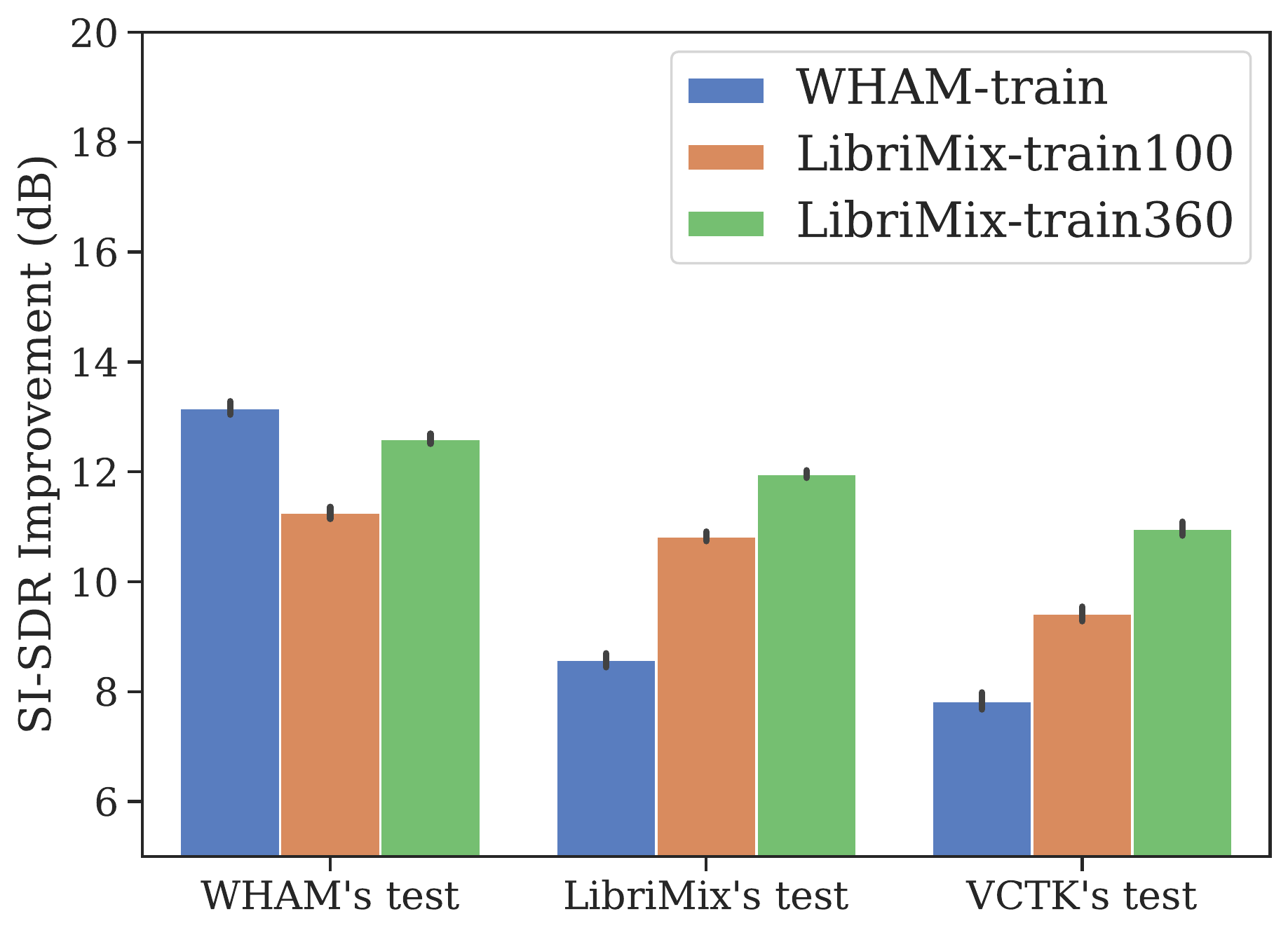}
    \caption{Cross dataset evaluation on the noisy separation task. Errors bars indicate 95 \% confidence interval}
    \label{fig:sep_noisy}
\end{figure}

The experiments in \cite{EmpiricalDolby2020} have shown that models trained on wsj0-2mix do not generalize well to other datasets. Similarly to \cite{EmpiricalDolby2020}, we investigate the generalization ability of Conv-TasNet when trained on different datasets. We train six different Conv-TasNet models on  WHAM!  \texttt{train}, LibriMix  \texttt{train-100} and \texttt{train-360} in both clean and noisy condition. We evaluate each model on the corresponding (clean or noisy) test sets of Libri2Mix, WHAM!, and VCTK2Mix. The results in clean and noisy conditions are shown in Figs.~\ref{fig:sep_clean} and \ref{fig:sep_noisy}, respectively. Note that noise samples are matched across the three noisy test sets. For both clean and noisy separation, we can see that WHAM!-trained models poorly generalize to LibriMix, with a 4~dB SI-SDR drop compared to LibriMix-trained models, while LibriMix-trained models obtain closer performance to WHAM!-trained models on the WHAM! test set with a 0.8~dB SI-SDR drop only. On the clean and noisy versions of VCTK-2mix, WHAM!-trained models perform around 3--4~dB less well than models trained on Librimix's \texttt{train-360}. The general performance drop from models trained on LibriMix's \texttt{train-100} compared to LibriMix's \texttt{train-360} confirms again that the amount of data is key to better generalization and that the amount of data available in WHAM! is insufficient.
Altogether, these results indicate that the clean and noisy versions of LibriMix allow better generalization than the wsj0-mix and WHAM! datasets. 

Several factors can influence generalization. While VCTK-2mix was generated with statistics matching the ones in LibriMix, we argue that this is not the reason, as results reported in  \cite{EmpiricalDolby2020} go in the same direction. Instead, we believe that the number of speakers (100 against 900), the size of the vocabulary (5k against 60k), the recording conditions (same room same recording material against varying rooms and material) and the total amount of training data (30~h against 212~h) add up to explain that models trained with LibriMix's \texttt{train-360} offer better generalization.

Results reported in \cite{EmpiricalDolby2020} are somewhat different than the ones we report here, which can be explained by several factors.
First, the VCTK-based two-speaker test set in \cite{EmpiricalDolby2020} was designed using the Matlab scripts from \cite{DPCLHershey2016}. These scripts do not remove silences and compute SNRs based on signal power instead of LUFS. As utterances from VCTK can be filled with silence, this greatly increases the effective SNR range of mixtures. For example, a short utterance in a long silence mixed at 0~dB with a long utterance without silence can produce a mixture where the second source is almost inaudible. This explains the low performance obtained on VCTK in \cite{EmpiricalDolby2020}.
Second, the alternative training and test sets are based on LibriTTS \cite{LibriTTS2019} which is itself derived from LibriSpeech \cite{panayotov2015librispeech}. LibriTTS has shorter and cleaner utterances, which could explain the higher performance reported on its test set in \cite{EmpiricalDolby2020}, and the larger drop in performance when tested on wsj0-2mix's test set. 

\section{Conclusions} \label{sec:conclusions}
In this work, we introduced LibriMix, a new family of datasets for generalizable single-channel speech separation. Libri2Mix and Libri3Mix enable two- and three-speaker separation in clean and noisy conditions. We report competitive results in all conditions using Asteroid's implementation of Conv-TasNet.
A new independent test set, VCTK-2mix, is also released to enable reproducible cross-dataset evaluation. Experiments show that models trained on Libri2Mix generalize better to VCTK-2mix than models trained with WHAM!. Additionaly, Libri3Mix is the first open-source dataset to enable three-speaker noisy separation.
Stepping towards more realistic scenarios, we release SparseLibri2Mix and SparseLibri3Mix, two- and three-speaker test sets consisting of sparsely overlapping speech mixtures with a varying amount of overlap. Initial results reported on it suggest that there still is room for improvement on this scenario.
Future work includes the design of a training set of sparsely overlapping speech mixtures, as well as a more diverse set of noise samples.

\bibliographystyle{IEEEtran}

\bibliography{mybib}

\end{document}